\documentclass[conference]{IEEEtran}
\IEEEoverridecommandlockouts
\makeatletter
\renewcommand{\fnum@figure}{Fig. \thefigure}
\makeatother
\usepackage[noadjust]{cite}
\usepackage{filecontents}
\usepackage{balance}

\usepackage{amsfonts}
\usepackage[pdftex]{graphicx}
\graphicspath{{../pdf/}{../jpeg/}}
\DeclareGraphicsExtensions{.pdf,.jpeg,.png}
\usepackage[cmex10]{amsmath}
\usepackage{mathabx}
\usepackage{algorithmic}
\usepackage{array}
\usepackage{mdwmath}
\usepackage{mdwtab}
\usepackage{eqparbox}
\usepackage{url}
\usepackage{color,soul}
\usepackage{amsmath} 
\usepackage{algorithm} 
\usepackage{multirow} 
\usepackage{dirtytalk} 

\usepackage[utf8]{inputenc}
\usepackage[english]{babel}
\usepackage{fancyhdr}
\usepackage{lastpage}
\usepackage{caption}
\usepackage{subcaption}
\usepackage{adjustbox}
\usepackage{amssymb}

\usepackage{xcolor}
       
\usepackage{calrsfs}
\DeclareMathAlphabet{\pazocal}{OMS}{zplm}{m}{n}
\newcommand{\La}{\mathcal{L}}

\usepackage[acronym]{glossaries}
\newacronym{OFDM}{OFDM}{orthogonal frequency division multiplexing}
\newacronym{CP}{CP}{cyclic prefix}
\newacronym{JCAS}{JCAS}{joint communication and  radar sensing}
\newacronym{RF}{RF}{radio frequency}
\newacronym{QAM}{QAM}{quadrature amplitude modulation}
\newacronym{MUSIC}{MUSIC}{multiple signal classification}
\newacronym{ESPRIT}{ESPRIT}{estimation of signal parameters via rotational invariance technique}
\newacronym{CS}{CS}{compressive sensing}
\newacronym{FPS}{FPS}{Fourier projection-slice}
\newacronym{LMMSE}{LMMSE}{linear minimum mean square error}
\newacronym{LS}{LS}{least square }
\newacronym{MMSE}{MMSE}{minimum mean square error}
\newacronym{CE}{CE}{channel estimation}
\newacronym{RMSE}{RMSE}{root mean square error}
\newacronym{TDD}{TDD}{Time Division Duplexing}
\newacronym{FDD}{FDD}{Frequency Division Duplexing}
\newacronym{UL}{UL}{Uplink}
\newacronym{DL}{DL}{deep learning}
\newacronym{BS}{BS}{Base Station}
\newacronym{MRC}{MRC}{Maximum Ratio Combining}
\newacronym{CSI}{CSI}{Channel State Information}
\newacronym{ACI}{ACI}{adjacent channel interference}
\newacronym[longplural={User Equipments}]{UE}{UE}{User Equipment}
\newacronym{3GPP}{3GPP}{3rd generation partnership project }
\newacronym{LTE}{LTE}{long term evolution}
\newacronym{GT}{GT}{ground-truth}
\newacronym{SFT}{SFT}{sparse Fourier transform}
\newacronym{ED}{ED}{Energy Detection}
\newacronym{ASK}{ASK}{Amplitude-shift keying}
\newacronym{PSK}{PSK}{Phase-shift keying}
\newacronym{ZCP}{ZCP}{Zadoff-Chu precoding}
\newacronym{ISM}{ISM}{industrial scientific and medical}
\newacronym{SNR}{SNR}{signal-to-noise ratio}
\newacronym{IDFT}{IDFT}{inverse discrete Fourier transform }
\newacronym{DFT}{DFT}{discrete Fourier transform }
\newacronym{IFFT}{IFFT}{inverse fast Fourier transform }
\newacronym{FFT}{FFT}{fast Fourier transform }
\newacronym{ISI}{ISI}{inter symbol interference}
\newacronym{ICI}{ICI}{inter carrier interference}
\newacronym{MIMO}{MIMO}{Multiple Input Multiple Output}
\newacronym{PAPR}{PAPR}{peak-to-average power ratio}
\newacronym{GLRT}{GLRT}{Generalized Likelihood Ratio Test}
\newacronym{i.i.d}{i.i.d}{independent and identically distributed}
\newacronym{PSNR}{PSNR}{peak signal to noise ratio}
\newacronym{AWGN}{AWGN}{additive white Gaussian noise}
\newacronym{ZCT}{ZCT}{Zaddof-Chu transform}
\newacronym{HPA}{HPA}{high power amplifier}
\newacronym{RFI}{RFI}{Radio Frequency Interferences}
\newacronym{2D}{2D}{two-dimensional}
\newacronym{1D}{1D}{one-dimensional}
\newacronym{MSE}{MSE}{mean square error}
\newacronym{BER}{BER}{bit error rate}
\newacronym{DNN}{DNN}{deep neural network}
\newacronym{CNN}{CNN}{convolutional neural network}
\newacronym{RCS}{RCS}{radar cross section}
\newacronym{ITS}{ITS}{Intelligent Transportation Systems}
\newacronym{ReLU}{ReLU}{rectified linear unit}

\def\BibTeX{{\rm B\kern-.05em{\sc i\kern-.025em b}\kern-.08em
    T\kern-.1667em\lower.7ex\hbox{E}\kern-.125emX}}
\begin{document}

\bstctlcite{IEEEexample:BSTcontrol}
\title{Deep Learning-based Estimation for Multitarget Radar Detection\\
\thanks{This work was sponsored by the Junior Faculty Development program under the UM6P-EPFL Excellence in Africa Initiative.}
}

\author{\IEEEauthorblockN{Mamady  Delamou\IEEEauthorrefmark{1},
 Ahmad Bazzi\IEEEauthorrefmark{2}, Marwa Chafii\IEEEauthorrefmark{2}\IEEEauthorrefmark{3}, El Mehdi Amhoud\IEEEauthorrefmark{1}
}
\IEEEauthorblockA{\IEEEauthorrefmark{1}School of Computer Science, Mohammed VI Polytechnic University, Ben Guerir, Morocco,\\ \IEEEauthorrefmark{2} Engineering Division New York University (NYU) Abu Dhabi, UAE 
\\ \IEEEauthorrefmark{3} NYU WIRELESS, NYU Tandon School of Engineering, Brooklyn, NY
} 
{\{mamady.delamou, elmehdi.amhoud\}@um6p.ma, \{ahmad.bazzi, marwa.chafii\}@nyu.edu}}

\maketitle
\begin{abstract}
Target detection and recognition is a very challenging task in a wireless environment where a multitude of objects are located, whether to effectively determine their positions or to identify them and predict their moves. In this work, we propose a new method based on a convolutional neural network  (CNN) to estimate the range and velocity of moving targets directly from the range-Doppler map of the detected signals. We compare the obtained results to the two dimensional (2D) periodogram, and to the similar state of the art methods, 2DResFreq and \mbox{VGG-19} network and show that the estimation process performed with our model provides better estimation accuracy of range and velocity index in different signal to noise ratio (SNR) regimes along with a reduced prediction time. Afterwards, we assess the performance of our proposed algorithm using the peak signal to noise ratio (PSNR) which is a relevant metric to analyse the quality of an output image obtained from compression or noise reduction. Compared to the 2D-periodogram,  2DResFreq and VGG-19, we gain 33 dB, 21 dB and 10 dB, respectively, in terms of PSNR when \mbox{SNR = 30 dB}.
\end{abstract}
\begin{IEEEkeywords}
Convolutional neural network, joint communication and sensing, monostatic radar
\end{IEEEkeywords}
\section{Introduction}
Network densification is one of the prominent building blocks for future wireless communication systems. In addition to high cell density, dense networks will integrate a large amount of connected drones and autonomous vehicles. Hence, setting up an effective object detection system becomes an important challenge for the wireless infrastructure. Object detection is a topic that is addressed from different angles as it becomes the focus of future technologies. It is investigated in joint communication and radar systems (JCRS) \cite{JCASsurvey,zhang2020perceptive,10061453,10018908} to improve radar detection in a merged communication and detection system.\\ 
\indent It is important to note that most parameter detection problems are complex because of their non-linearity \cite{JCASsurvey}. Therefore, much effort is  put into either proposing new algorithms or refining existing solutions, such as removing or reducing side lobes around the solutions, or even improving the robustness of the algorithms in low signal to noise ratio (SNR) regions. Among all target detection algorithms, the periodogram technique which is based on the discrete Fourier transform (DFT) was widely investigated \cite{braun2014ofdm}. Moreover, compressive sensing takes advantage of the sparsity property in some signals to reduce the number of samples needed for estimation \cite{chandrakanth2016compressed,anitori2013compressive}. However, it is subject to degradation of image resolution. Furthermore, the matrix pencil \cite{hua1992estimatingmatrixpencil} is an algebraic method that solves the parameter estimation problem by approximating a function by a sum of complex exponentials.\\
\indent In addition, the multiple signal classification (MUSIC) and the estimation of signal parameters by rotational invariance (ESPRIT) methods are two well-known parametric estimators \cite{roy1987comparativemusicesprit}.
They can reach super-resolutions, however, they require very large number of samples \cite{wu2017fastmusicesprit,liu2020supermusicesprit}. Moreover, it is known that one of the main limitations of the \gls{2D} matrix pencil and ESPRIT  is the matching of the frequency pairs. Matrix pencil matching could be done as proposed in Step 3 of Subalgorithm 1 in \cite{bazzi2016singlematrixpencil}. Finally, MUSIC and ESPRIT  have a high computational complexity, and as the SNR decreases, their performance  degrades rapidly \cite{pan2021deep2DResFreq}.\\
\indent Furthermore, \gls{DL}-based algorithms for solving complex problems in many areas, including radar signal processing have gained considerable attention in recent years \cite{geng2021deep}. This is due to the excellent ability of \gls{DL} models to extract miscellaneous features that classical techniques fail in doing. Recently, Pingping et al., put forward 2DResFreq in \cite{pan2021deep2DResFreq}, which is based on \gls{DL} and aims at extracting several sinusoids in a \gls{2D} signal. This is an extension of the work proposed in \cite{izacard2019data} from a \gls{1D} signal to a \gls{2D} one. The work initiated in \cite{simonyan2014very} is a convolutional neural network model called VGG-19, which can achieve very good accuracy with the ImageNet dataset. In \cite{yavuz2020radar}, the proposed technique comprises a \gls{CNN} for target detection with a typical pulsed Doppler radar. The neural network generates range-Doppler data for only a single target with an isotropic antenna which is not practical in a dense network. In addition, in \cite{carrera2020target}, the authors came up with a machine learning model for processing echoed signals to determine whether a valid target exists. The model performs target detection based on random decision forests and recurrent neural networks, but don't take into account the range and velocity of those targets.\\
\indent In this work, we propose a new \gls{DL} framework for learning the radar range-Doppler map. The latter is a \gls{2D} representation of the time-Doppler shift couple, which is translated into range-velocity information. Compared to the work presented in \cite{pan2021deep2DResFreq}, instead of estimating the 2D-frequencies, we learn the range-Doppler map which is mainly converted into an image processing task. Therefore, the model matches each estimated channel to its corresponding range-Doppler map.\\
\indent Despite that many image classification and recognition tasks have benefited from \gls{CNN}, recent evidence reveals that network depth is of crucial importance \cite{simonyan2014very}. The question to be answered is if learning better networks is as easy as stacking more layers. As mentioned in \cite{he2016deep}, with the increase of network depth, accuracy gets saturated and then degrades rapidly. Such degradation is not caused by overfitting, and adding more layers to a suitable \gls{DL} model leads to higher training error. As reported in \cite{he2015convolutional} and thoroughly verified in \cite{he2016deep}, the degradation can be mitigated by introducing a deep residual learning framework with shortcut connections. With this is mind, our contributions are summarized as follows: (i) introducing a \gls{CNN} tailored for radar range-Doppler estimation, which is then trained on synthetic data and
(ii) testing the \gls{CNN} on newly generated signals, and comparing with other state of the art methods, i.e., 2DResFreq and VGG-19. We  report an improved range and velocity \gls{RMSE}, a high noise reduction, along with high \gls{PSNR} and a lower prediction time.
\newline
\indent The remainder of the paper is organized as follows: In Section II, we introduce the system model and formulate the radar detection problem. In Section III, we detail the structure of our proposed \gls{CNN} model. In Section IV, we present simulation results. Finally, in Section V, we conclude and set forth our perspectives.
\section{System model and problem statement}
\begin{figure}[t!] 
\centering
\includegraphics[width=3.5in]{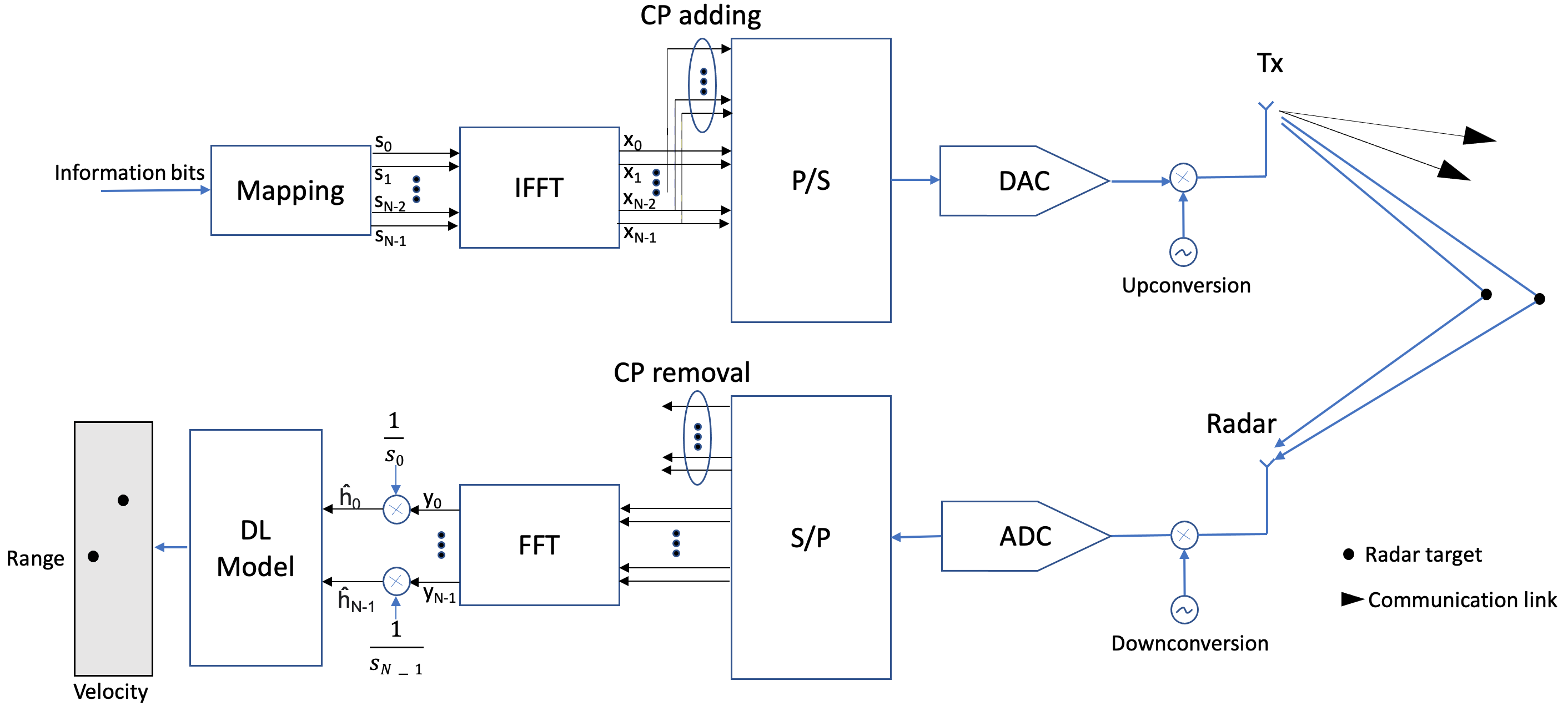}
\caption{A mono-static co-located integrated radar and communication system, where the signal used for communication is then re-used for radar processing.}
\label{systemmodel}
\end{figure}

We consider a wireless communication system consisting in a communication transmitter co-located with a monostatic radar. The transmitted signal from the communication antenna is perfectly known to the radar, and is reflected by targets characterized by their range and velocity as illustrated in \mbox{Fig. \ref{systemmodel}}. We assume that the interference between the reflected signal (radar signal) and the communication signal is perfectly managed. The transmitter and receiver exchange \gls{OFDM} frames.  The total bandwidth $B$ is divided into $N$ small bands with central frequencies $f_0$,$f_1$\dots$f_{N-1}$ such that \mbox{$\Delta f=\frac{B}{N}$}. The OFDM symbol duration $T$ is given by \mbox{$T=\frac{1}{\Delta f}$}, where $\Delta f$ is the OFDM frequency spacing.\\
\indent We consider that an \gls{OFDM} frame composed of $M$ \gls{OFDM} symbols, with \mbox{$\mathbf{S}$} representing the transmitted \gls{QAM} symbols matrix, \mbox{$\mathbf{X}$} is the \gls{OFDM} frame and \mbox{$\mathbf{H}$} the channel matrix. The matrices $\mathbf{S}$, and $\mathbf{X}$ have the same dimension, i.e., $N$ rows and $M$ columns. Since a \gls{CP} is added between consecutive symbols within the frame, the number of \gls{QAM} symbols in each \gls{OFDM} symbol becomes $N_s=N+N_{cp}$, where $N_{cp}$ is the number of \gls{QAM} symbols transmitted in \gls{CP} duration. The total OFDM symbol transmission time $T_s$ becomes \mbox{$T_s = T+T_{cp}$}, with $T_{cp}$ the \gls{CP} duration. The conversion from digital to analog is accomplished within a dedicated digital-to-analogic converter (DAC) and the signal is up-converted using the carrier frequency $f_c$. At the radar, the~\gls{CP} is removed, and then \gls{FFT} is performed on the~\gls{OFDM} bandpass signals. Finally, after the spectral division, targets detection algorithm is applied. The received signal can be written as $\mathbf{Y=S \cdot  H+Z}$, where $\mathbf{Z}$ is the noise matrix.\\
\indent By considering the baseband signal $x(t)$, the transmitted passband  signal is \mbox{$x_{pb}(t)=x(t)e^{j2\pi f_ct}$}. For a target $p$ at distance $d_p$ from the transmitter, and moving at velocity $v_p$, the received passband signal at the radar is impacted by the following effects \mbox{\cite{braun2014ofdm}:}
 \begin{itemize}
     \item The attenuation factor $b_p$ which depends on the distance $d_p$, the radar cross section (RCS) $\sigma_{RCS}$, the carrier frequency $f_c$ and the speed of light $c$. By using Friis equation of transmission, we have $b_p=\sqrt{\frac{c\sigma_{RCS}}{(4\pi)^3d_p^4f_c^2}}.$

     \item The signal delay $\tau_p$ caused by the round-trip, $\tau_p=\frac{d_p}{c}$.
     \item The Doppler-shift $f_{D_p}$ caused by the velocity $v_p$ of the target, $f_{D_p}=\frac{2v_p}{c}f_c$.
     \item The random rotation phase $\varphi$ introduced when the signal hits the target.
     \item The additive white Gaussian noise (AWGN) $z(t)$ such that \( z(t) \sim \mathcal{N}(\mu,\,\sigma^{2}) \).
 \end{itemize}
By denoting $N_t$, the total number of moving targets. The estimated channel  $\mathbf{\hat{H}}$ has entries given by  \cite{braun2014ofdm,zerhouni-access,delamouefficient}:
\begin{equation}\label{ratio}
\begin{aligned}
\hat{h}_{k, l}=\frac{y_{k, l}}{s_{k, l}}=\sum_{p=0}^{N_{t}-1} b_{p} e^{j 2 \pi \frac{l N_{s} f_{D_{p}}}{N \Delta_{f}}} e^{-j 2 \pi k \Delta f \tau_{p}} e^{j \Phi}+\tilde{z}_{k, l},\\
\text{with}~ 0 \leqslant k \leqslant N-1, 0 \leqslant l \leqslant M-1,
\end{aligned}
\end{equation}\label{spectral_div}
where $s_{k, l}$, $y_{k, l}$, and $\hat{h}_{k, l}$ are the ($k$,$l$)th entry of $\boldsymbol{S}$, $\boldsymbol{Y}$, $\mathbf{\hat{H}}$, respectively. $\Phi$, $\tilde{z}_{k, l}$ represent the phase added after reflection and the ($k$,$l$)th entry of the noise matrix $\tilde{\mathbf{Z}}$, obtained after the spectrum division, respectively.\\
Let us write $f_{p,1}$ and  $f_{p,2}$ as
\begin{equation}\label{express fp}
   f_{p,1}=\Delta f \times \tau_p, ~\mathrm{and}~ f_{p,2}=T_s\times f_{D_p},
\end{equation}
with $f_{D_p}=\frac{2v_p}{c}f_c$, $\tau_p=\frac{2d_p}{c}$, and $T_s=\frac{N_s}{N \Delta f}$. From (\ref{express fp}), (\ref{ratio}) can be written as
\begin{equation}\label{ratio2}
\begin{aligned}
\hat{h}_{k, l}=\sum_{p=0}^{N_t-1} b_{p} \exp \left(-j 2 \pi f_{p, 1} k\right) \exp \left(j 2 \pi f_{p, 2} l\right) e^{j \Phi}+\tilde{z}_{k, l},\\
\text{with}~ 0 \leqslant k \leqslant N-1, 0 \leqslant l \leqslant M-1.
\end{aligned}
\end{equation}
\indent The target detection problem consists in estimating $f_{D_p}$, $\tau _p$ and $b_p$ from which we retrieve range, velocity and the reflectance of targets. Hence, we turn the problem into estimating $f_{p, 1}$, $f_{p, 2}$ and $b_{p}$. This is equivalent to estimating the couples of indexes ($\hat{k}_p$, $\hat{l}_p$) corresponding to the indexes of ($f_{p, 1}$, $f_{p, 2}$) in $\mathbf{\hat{H}}$. $\hat{k}_p$ and $\hat{l}_p$ are referenced as range and velocity indexes, respectively. Once we get the list of the indexes \mbox{($\hat{k}_p$, $\hat{l}_p$)}, the corresponding \mbox{($f_{p, 1}$, $f_{p, 2}$)} are deduced. Finally ranges and velocities can be retrieved from (\ref{express fp}).
\section{Deep learning-based multitarget detection}
\subsection{\gls{DL} architecture}
\begin{figure*}[ht]
     \centering
     \begin{subfigure}[b]{0.75\textwidth}
         \centering
         \includegraphics[width=\textwidth,height=4cm]{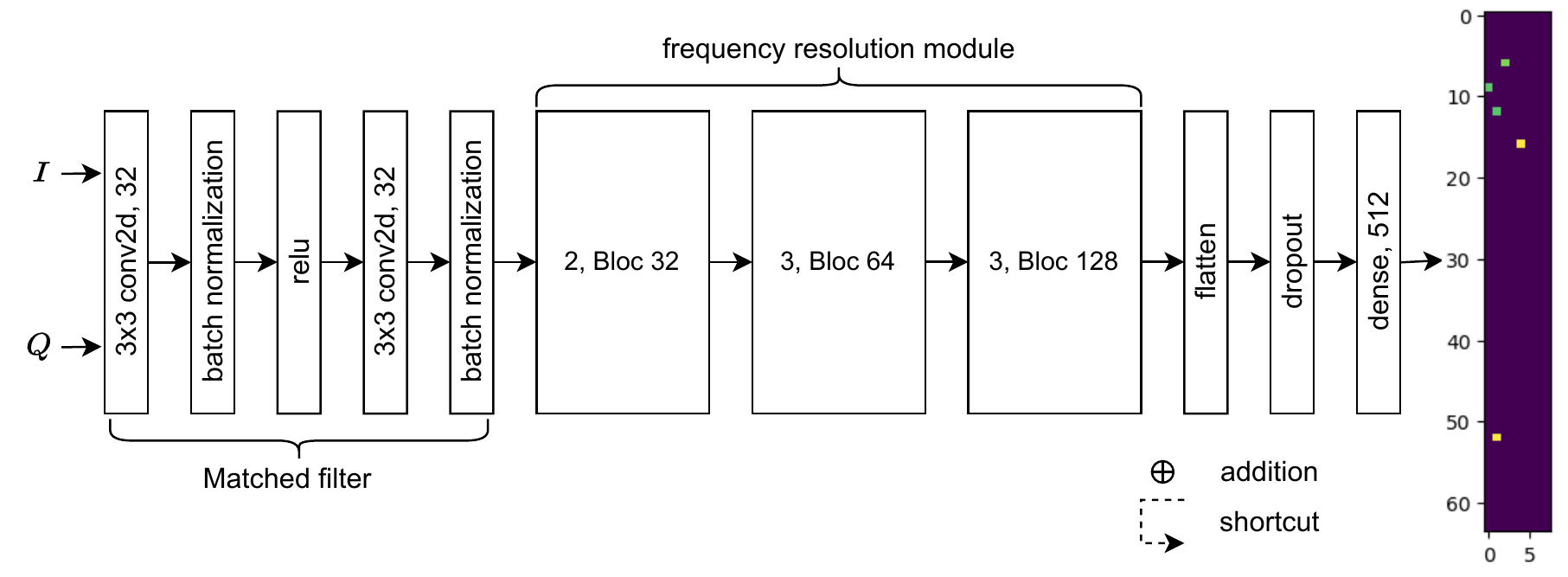}
         \caption{\gls{CNN} structure taking as input $I$ and $Q$ and output the range-Doppler map. The frequency resolution module is composed of three blocs. The description of a bloc is shown in Fig. \ref{shortcut_structure}.}
         \label{main_dnn_structure}
     \end{subfigure}
     \hfill
     \begin{subfigure}[b]{0.2\textwidth}
         \centering
         \includegraphics[width=\textwidth,height=4cm]{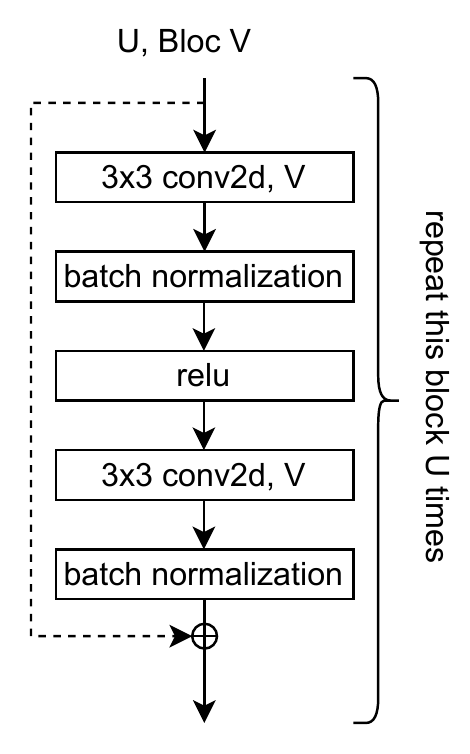}
         \caption{Bloc structure with $U$ the number of block repetition and $V$ the filter size of the convolutional layer.}
         \label{shortcut_structure}
     \end{subfigure}
     \caption{Data structure, convergence and sample prediction}
        \label{dnn_architecture}
\end{figure*}
In this section, we present our \gls{DL} model used to estimate the targets range and velocity index. The optimization problem introduced consists in estimating the dominant frequencies' index contained in $\mathbf{\hat{H}}$, which can be found in the radar range-Doppler map. Some numerical approaches can be introduced to solve it. Instead, we introduce \gls{DL} because of  its ability to extract miscellaneous features that are hidden from classical techniques. The \gls{DL} first reduces the noise by filtering the noisy signal, and greatly learns the main features contained in the data, i.e., the peaks in this case.\\
\indent The input layer takes $\mathbf{(I, Q)}$ where $\mathbf{\hat{H}=I+jQ}$. The overall network is composed of convolution layers, batch normalization layers, \gls{ReLU}, dropout layers, and a dense layer. As shown in Fig. \ref{dnn_architecture}, the \gls{DL} network is based on deep residual learning. In fact, as mentioned in \cite{he2016deep}, as the depth of a network increases, the accuracy reaches saturation and at this point, adding more layers increases the training error. This behavior can be mitigated using the residual network with shortcut connections  \cite{he2015convolutional}. Based on this alternative, we achieve a deep learning model without saturation. However, adding a very large number of layers increases the complexity. Hence, to avoid heavy models, we are also limited by the training and the prediction time.

The \gls{CNN} we propose is composed of a matched filter whose output is such that it maximizes the ratio of output peak power to mean noise power. Moreover, to achieve good radar resolution, we must get better frequency resolution which refers to the minimum frequency difference below which two frequencies can not be distinguished. The residual shortcut connections are used throughout the frequency resolution module to achieve much deeper convergent model which improves the frequency resolution and consequently improves the radar resolution (Fig. \ref{dnn_architecture}).
\subsection{Model label generation}
As a model to be trained, the \gls{GT} label is crucial, it is the real range-Doppler map that the model matches the extracted features with, during a supervised learning. At the training stage, it is used to associate $\mathbf{\hat{H}}$ with its range-Doppler. Each target $p$ is characterized by a pair of frequencies ($f_{p,1}$,$f_{p,2}$) and the amplitude $b_p$ in accordance with (\ref{ratio2}). A \gls{GT}   range-Doppler map $g_t$ of dimension ($N$, $M$) contains zeros in all the matrix points except at entries where targets are located (ideal map). Let us consider three sets $F_1$, $F_2$ and $B$ such that $f_{r,1} \in F_1$, $f_{s,2} \in F_2$ and $b \in B$ for any $r \in \{0;1;2;...;N_{F_1}-1\}$ and for any $s \in \{0;1;2;...;N_{F_2}-1\}$. $N_{F_1}$ and $N_{F_2}$ are the cardinalities of $F_1$ and $F_2$, respectively. The \gls{GT}   range-Doppler map is constructed as follows:\\
\begin{itemize}
    \item For each target $p$, select randomly $f_{p,1} \in F_1$ and \mbox{$f_{p,2} \in F_2$} with $i_p$, $j_p$ the indexes of $f_{p,1}$ and $f_{p,2}$ in $F_1$ and $F_2$,  respectively.
    \item The indexes of $f_{p,1}$ and $f_{p,2}$ in the \gls{GT} are \mbox{$k_p=\frac{i_p\times N}{N_{F_1}}$}, and \mbox{$l_p=\frac{j_p\times M}{N_{F_2}}$}, respectively.
    \item The \gls{GT} range-Doppler map is defined as
    \begin{equation}
     g_t(k_p,l_p)= \begin{cases}\beta \ln(\gamma b_p +1), & 0\le p \le N_t \\ 0, & \text { otherwise }\end{cases},
     \end{equation}
where $\beta$ and $\alpha$ are two constants. $\beta \ln(\gamma b_p +1)$ is the logarithm information of the amplitude $b_p$. It expresses the amplitude information of frequencies.
\end{itemize}

The loss function based on square error is given by:
\begin{equation}\label{loss}
\text { $\La (g_t,\hat{g})$ }=\sum_{k=0}^{N-1} \sum_{l=0}^{M-1} \left(\hat{g}(k,l)-g_t\left(k, l\right)\right)^2,
\end{equation}
where the frequency representation $\hat{g}$ denotes the output of the \gls{CNN}. The objective is to minimize $\La (g_t,\hat{g})$ over iterations.
\section{Simulation results}
\begin{figure*}[ht]
     \centering
     \begin{subfigure}[b]{0.32\textwidth}
         \centering
         \includegraphics[width=\textwidth]{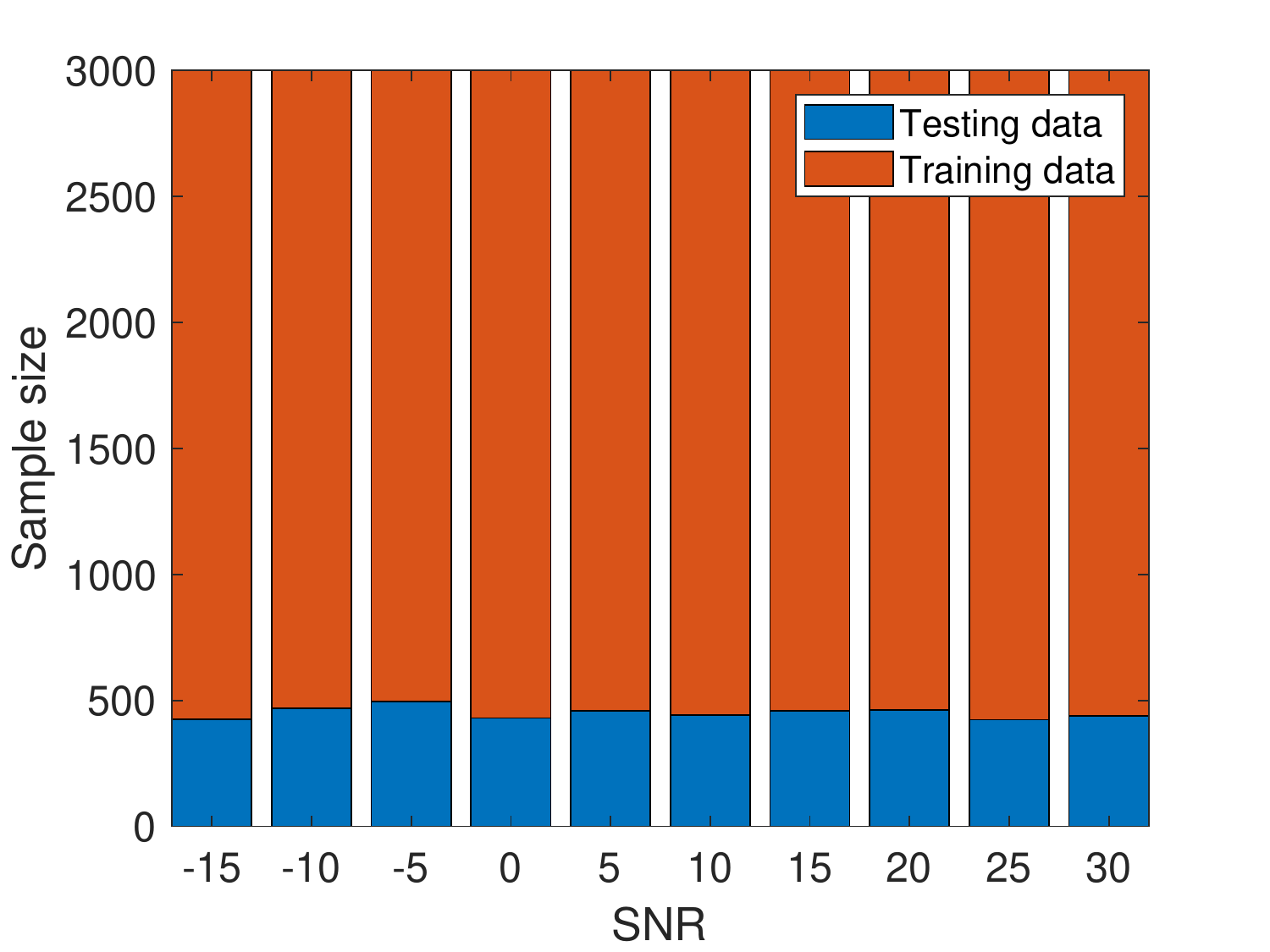}
         \caption{Sample distribution per SNR.}
         \label{data_distribution}
     \end{subfigure}
     \hfill
     \begin{subfigure}[b]{0.32\textwidth}
         \centering
         \includegraphics[width=\textwidth]{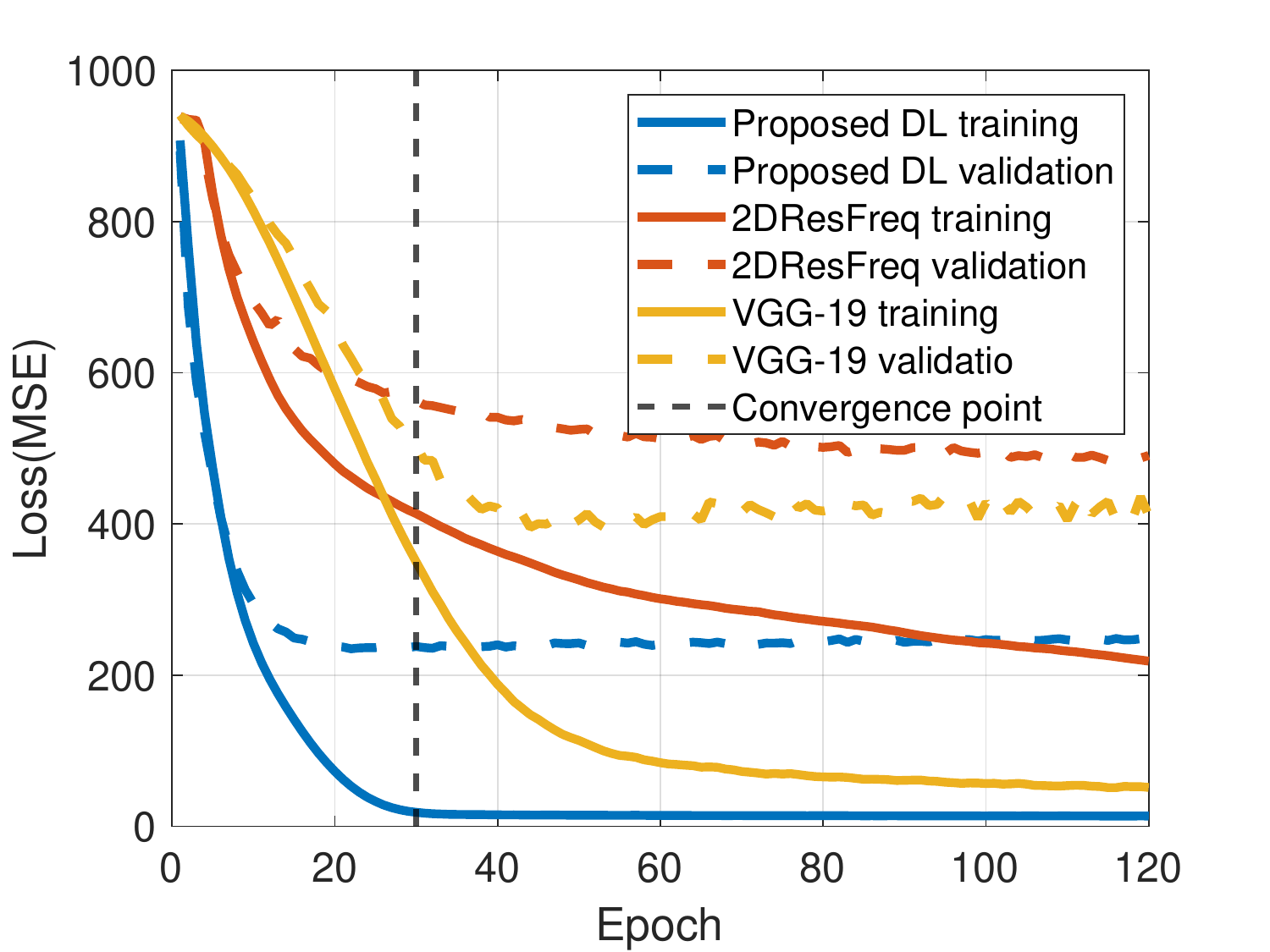}
         \caption{Model training and validation error.}
         \label{losses}
     \end{subfigure}
     \hfill
     \begin{subfigure}[b]{0.32\textwidth}
         \centering
         \includegraphics[width=\textwidth]{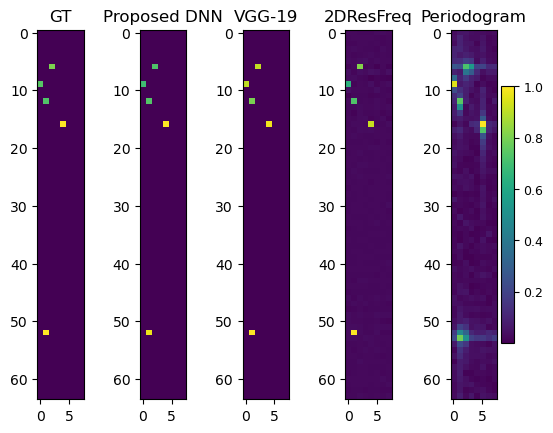}
         \caption{random range-Doppler map sample.}
         \label{examplespred}
     \end{subfigure}
        \caption{Data structure, convergence and sample prediction.}
        \label{}
\end{figure*}
\begin{figure*}[ht]
     \centering
     \begin{subfigure}[b]{0.32\textwidth}
         \centering
         \includegraphics[width=\textwidth]{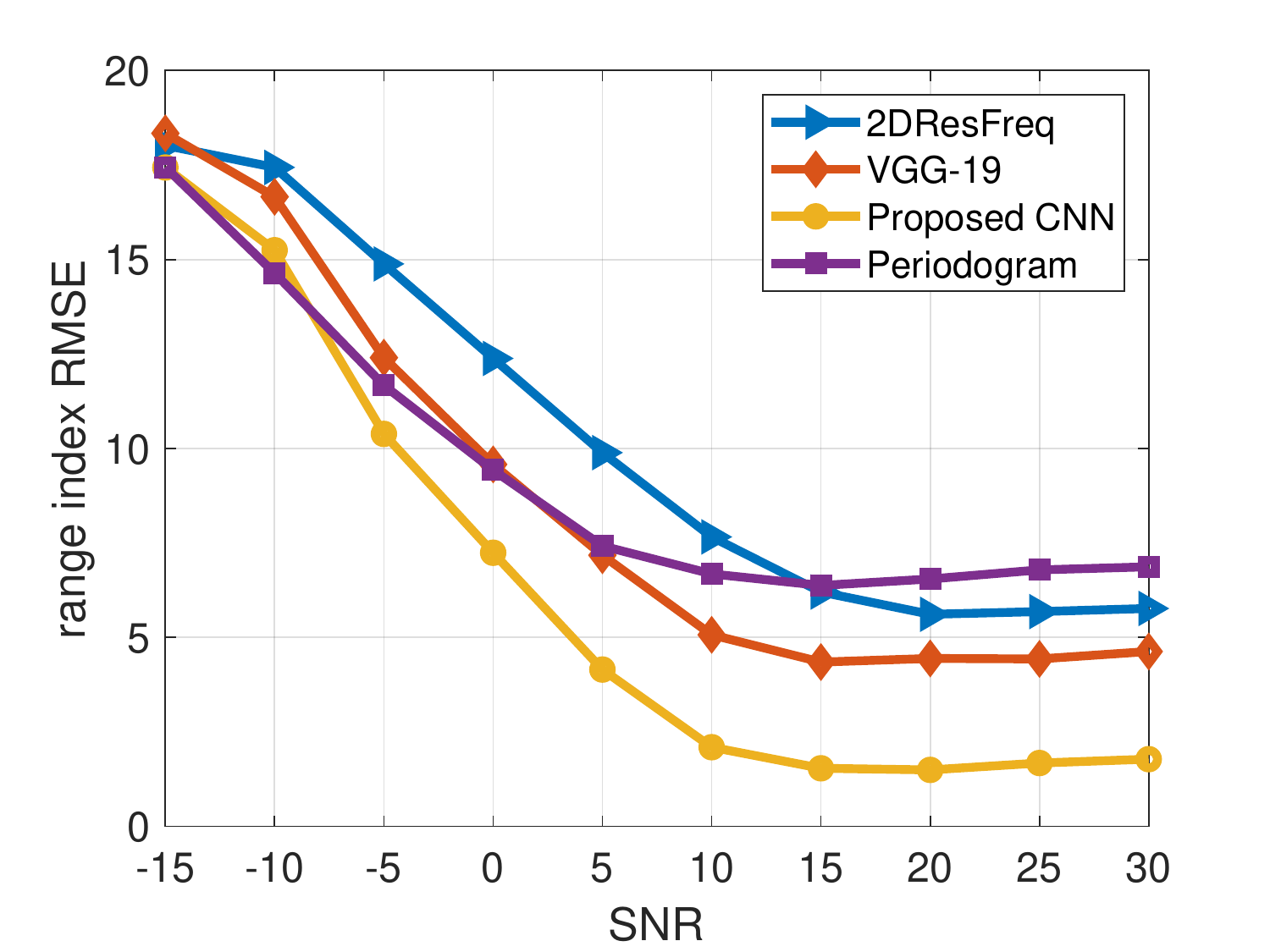}
         \caption{Range RMSE as a function of the SNR.}
         \label{range rmse}
     \end{subfigure}
     \hfill
     \begin{subfigure}[b]{0.32\textwidth}
         \centering
         \includegraphics[width=\textwidth]{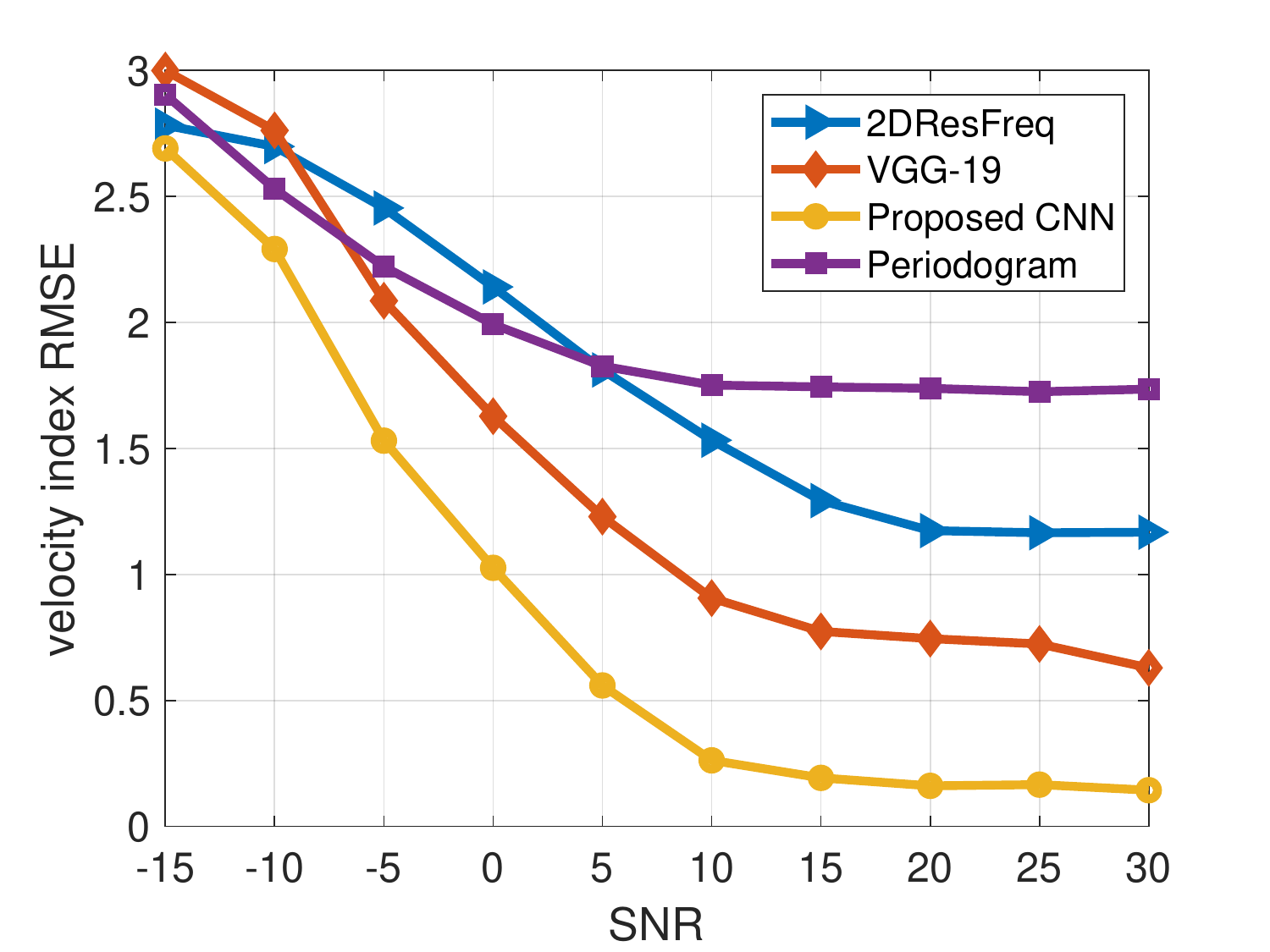}
         \caption{Velocity RMSE as a function of the SNR.}
         \label{velo rmse}
     \end{subfigure}
     \hfill
     \begin{subfigure}[b]{0.32\textwidth}
         \centering
         \includegraphics[width=\textwidth]{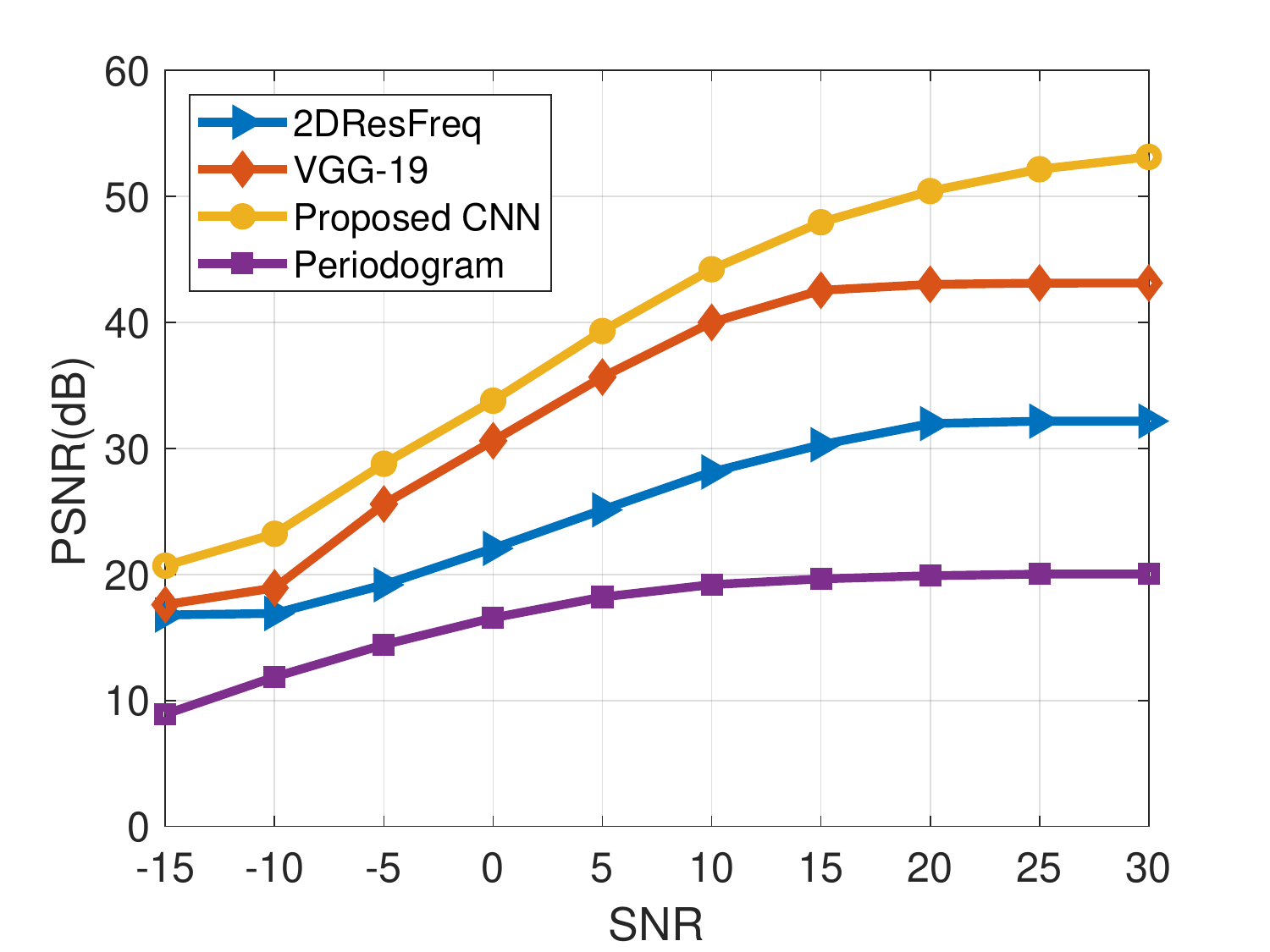}
         \caption{PSNR as a function of the SNR.}
         \label{pnsr}
     \end{subfigure}
        \caption{Comparison of the estimation of models}
        \label{psnr}
\end{figure*}
In this section, we assess the performance of our \gls{CNN} in terms of the \gls{RMSE} of the range and velocity index estimation, the \gls{PSNR} of the estimated range-Doppler map and the model training and prediction time. All the simulation has been done on a computer provided with an Intel Xeon CPU 2.20 GHz, 13 GB RAM, Tesla K80 gas pedal, and 12 GB GDDR5 VRAM. The learning rate is initialized to $8\times 10^{-5}$, the batch size is set to 64, the dropout factor to 0.5 and $\beta=\gamma=100$. The number of targets to be randomly predicted is $N_t$ = 5. For each given target $p$, the amplitude $b_p$ is the absolute value of $0.1+r_g$, and the phases are chosen to be the same, with $r_g$ being  sampled from a standard Gaussian distribution. The frequency coordinates of the $p$th target, $f_{p,1}$ and $f_{p,2}$, are both generated within the range $[-0.5,0.5[$. We have fixed the dimensions of the radar ratio signal equal to $N=64$ and $M=8$. To avoid targets to be too close to each other, the minimum separation between the coordinates of any two targets on the $f_1$ axis is $\frac{1}{3N}$, whereas the minimum separation between the coordinates of any two components on the $f_2$ axis is $\frac{1}{3M}$ \cite{pan2021deep2DResFreq}. 
First, we generate 3,000 noise-free signals and their \gls{GT} following the previous mentioned configuration. For each noise free signal, the corresponding noisy signals are generated with SNR within $[-15, 30]$ dB, all matching the same equivalent \gls{GT}. We end up with a dataset of 30,000 entries, well distributed over all the SNR ranges as shown in Fig.~\ref{data_distribution}.\\
\indent Afterwards, we introduce the \gls{PSNR}, which expresses the ratio between the maximum value of a signal and the power of distorting noise that affects the quality of its representation. In the following implementation, we are dealing with a standard \gls{2D} arrays. The mathematical representation of the \gls{PSNR} for the estimated label $\hat{g}$ is given by:\\
\begin{equation}\label{psnr-express}
PSNR=20 \log _{10}\left(\frac{\max\{\hat{g}\}}{\sqrt{MSE}}\right),
\end{equation}
where the MSE is expressed as
\begin{equation}
MSE=\frac{1}{NM} \sum_{k=0}^{N-1} \sum_{l=0}^{M-1}\|\hat{g}(k, l)-g_t(k, l)\|^{2}.
\end{equation}
Here, $\max\{\hat{g}\}$ is the maximum pixel value of the image. When the pixels are represented  using $A$ bits per sample, $\max\{\hat{g}\}$ is $2A-1$. For this implementation, labels are normalized between 0 and 1 during the PSNR computation process.\\
\indent Fig. \ref{losses} depicts the loss of the training and validation sets for our model, 2DResFreq and VGG-19. From the figure, we notice that our model starts converging after almost 30 epochs whereas 2DResFreq and VGG-19 are still learning. This rapid convergence is due to batch normalization and \gls{ReLU} units.\\
\indent Fig. \ref{examplespred} shows a random sample \gls{GT} map, the corresponding predicted map using our model, VGG-19, 2DResFreq, and the estimated one using \gls{2D} periodogram. As it can be remarked in the figure, the periodogram suffers from side lobes, which are removed in the \gls{DL} models.\\
\indent Fig. \ref{range rmse} and Fig. \ref{velo rmse} present the RMSE of range and velocity index estimation as a function of the SNR. Since each target $p$ in the estimated map is characterized by its frequencies $f_{p,1}$ and $f_{p,2}$  with $k_p$, $l_p$ their respective indices, in a parameterized OFDM system where $\Delta f$, $f_c$ and $T_s$ are defined and fixed, the ranges and the velocities can be directly computed using $k_p$ and $l_p$ as described in (\ref{express fp}). Then, the RMSE of the estimated ranges and velocities can be calculated. Instead, we directly computed the RMSE of the indexes. The proposed \gls{CNN} outperforms the other approaches not only in range estimation but also in velocity estimation. For example, at \mbox{SNR = 30 dB}, in range index estimation, we have a \mbox{log(2.8) dB}, \mbox{log(3.9) dB} and \mbox{log(5) dB} gain with reference to VGG-19, 2DResFreq  and \gls{2D} periodogram, respectively. Similarly, in velocity index estimation, we have a \mbox{log(0.45) dB}, \mbox{log(1.2) dB} and \mbox{log(1.6) dB}  gain with reference to VGG-19, 2DResFreq  and \gls{2D} periodogram, respectively.\\
\indent In Fig. \ref{pnsr}, we plot the \gls{PSNR} of the output of the proposed \gls{CNN} and we compare it with VGG-19, 2DResFreq and 2D periodogram. In fact, regarding all the predicted maps, the \gls{2D} periodogram output is the most corrupted by noise compared to the \gls{DL} models, which layers reduce the noise effect on the input signal. In fact, compared to VGG-19, 2DResFreq and 2D periodogram, we gain \mbox{10 dB}, \mbox{21 dB}  and \mbox{33 dB}, respectively.\\
\indent Furthermore, we compute the training and prediction times of the three models over all the training and testing datasets respectively. We report the results in Table \ref{time_comparison}. As clearly shown from the table, our model is slightly slower than the VGG-19 during the training step which can be performed offline in real applications. Nonetheless, during the prediction step which is the most critical for radar applications, our DL model has the fastest prediction time.
\begin{table}[t]
\caption{Training and prediction time}
\centering
\begin{tabular}{c c c c }
\hline\hline
Model & Training time (s) & Time/epoch (s) & Prediction time (s) \\ [0.5ex] 
\hline
2DResFreq&5884.634&49&320.976\\
VGG-19&$\mathbf{2358.974}$&$\mathbf{20}$&285.165\\
Proposed \gls{DL}&3267.084&27&$\mathbf{261.941}$\\
\hline
\end{tabular}
\label{time_comparison}
\end{table}
\section{Conclusion}
In this work, we proposed to estimate the range and velocity of moving targets by using a \gls{CNN} that predicts the range-Doppler map directly from the channel estimates. The simulation results show that our model performs better in terms of the estimation error compared to VGG-19, 2DResFreq and 2D periodogram. Moreover, our proposed model outputs the range and velocity estimates in a short prediction time and has a high ability to reduce the  noise effect on the range-Doppler map which leads to a better detection accuracy. In our future work, we aim to extend this model to dynamic OFDM waveforms and longer frames.
\balance
\bibliographystyle{IEEEtran}
\bibliography{biblio_traps_dynamics}
\vspace{12pt}
\end{document}